\documentclass[useAMS]{mn2e} 

\usepackage{latexsym,graphicx}

\usepackage{color}

\newcommand\kms{{\rm\,km\,s^{-1}}}
\newcommand\msun{\rm\,M_\odot}

\newcommand\lsun{\rm\,L_\odot}

\def\apgt{\ {\raise-.5ex\hbox{$\buildrel>\over\sim$}}\ }
\def\aplt{\ {\raise-.5ex\hbox{$\buildrel<\over\sim$}}\ }

\title[TYC\,8606-2025-1: a mild barium star]{TYC\,8606-2025-1: a mild barium star surrounded by the ejecta of a very late
thermal pulse}
\author[V. V.~Gvaramadze et al.] 
       {V. V.~Gvaramadze,$^{1,2}$\thanks{E-mail: vgvaram@mx.iki.rssi.ru} Yu. V.~Pakhomov,$^3$ A. Y.~Kniazev,$^{4,5,1}$ \\
       \newauthor T. A.~Ryabchikova,$^3$ N.~Langer,$^6$ L.~Fossati$^7$ and E. K.~Grebel$^8$ \\
        $^1$Sternberg Astronomical Institute, Lomonosov Moscow State University, Universitetskij Pr. 13, Moscow 119992, Russia\\
        $^2$Space Research Institute, Russian Academy of Sciences, Profsoyuznaya 84/32, 117997 Moscow, Russia \\
        $^3$Institute of Astronomy, Russian Academy of Sciences, Pyatnitskaya 48, 119017, Moscow, Russia \\
        $^4$South African Astronomical Observatory, PO Box 9, 7935 Observatory, Cape Town, South Africa \\
        $^5$Southern African Large Telescope Foundation, PO Box 9, 7935 Observatory, Cape Town, South Africa \\
        $^6$Argelander-Institut f\"ur Astronomie, Auf dem H\"ugel 71, 53121 Bonn, Germany \\    
        $^7$Space Research Institute, Austrian Academy of Sciences, Schmiedlstrasse 6, 8042 Graz, Austria \\  
        $^8$Astronomisches Rechen-Institut, Zentrum f\"ur Astronomie der Universit\"at Heidelberg, M\"onchhofstr. 12-14, 
        D-69120 Germany \\
        }
\begin{document}

\date{Accepted 2019 September 01. Received 2019 August 30; in original form 2019 August 15}


\maketitle

\label{firstpage}

\begin{abstract}
We report the discovery of a spiral-like nebula with the {\it Wide-field Infrared Survey Explorer}
({\it WISE}) and the results of optical spectroscopy of its associated star TYC\,8606-2025-1 with the 
Southern African Large Telescope (SALT). We find that TYC\,8606-2025-1 is a G8\,III star of 
$\approx3 \, \msun$, showing a carbon depletion by a factor of two and a nitrogen enhancement by a 
factor of three. We also derived an excess of s-process elements, most strongly for barium, which is
a factor of three overabundant, indicating that TYC\,8606-2025-1 is a mild barium star. We thereby 
add a new member to the small group of barium stars with circumstellar nebulae. Our radial velocity 
measurements indicate that TYC\,8606-2025-1 has an unseen binary companion. The advanced evolutionary 
stage of TYC\,8606-2025-1, together with the presence of a circumstellar nebula, implies an initial 
mass of the companion of also about $3 \, \msun$. We conclude that the infrared nebula, due to its 
spiral shape, and because it has no optical counterpart, was ejected by the companion as a consequence 
of a very late thermal pulse, during about one orbital rotation.
\end{abstract}

\begin{keywords}
stars: abundances -- circumstellar matter -- stars: individual: TYC\,8606-2025-1 -- stars: low-mass
\end{keywords}

\section{Introduction}
\label{sec:intro}

\begin{figure*}
\begin{center}
\includegraphics[width=16cm,angle=0]{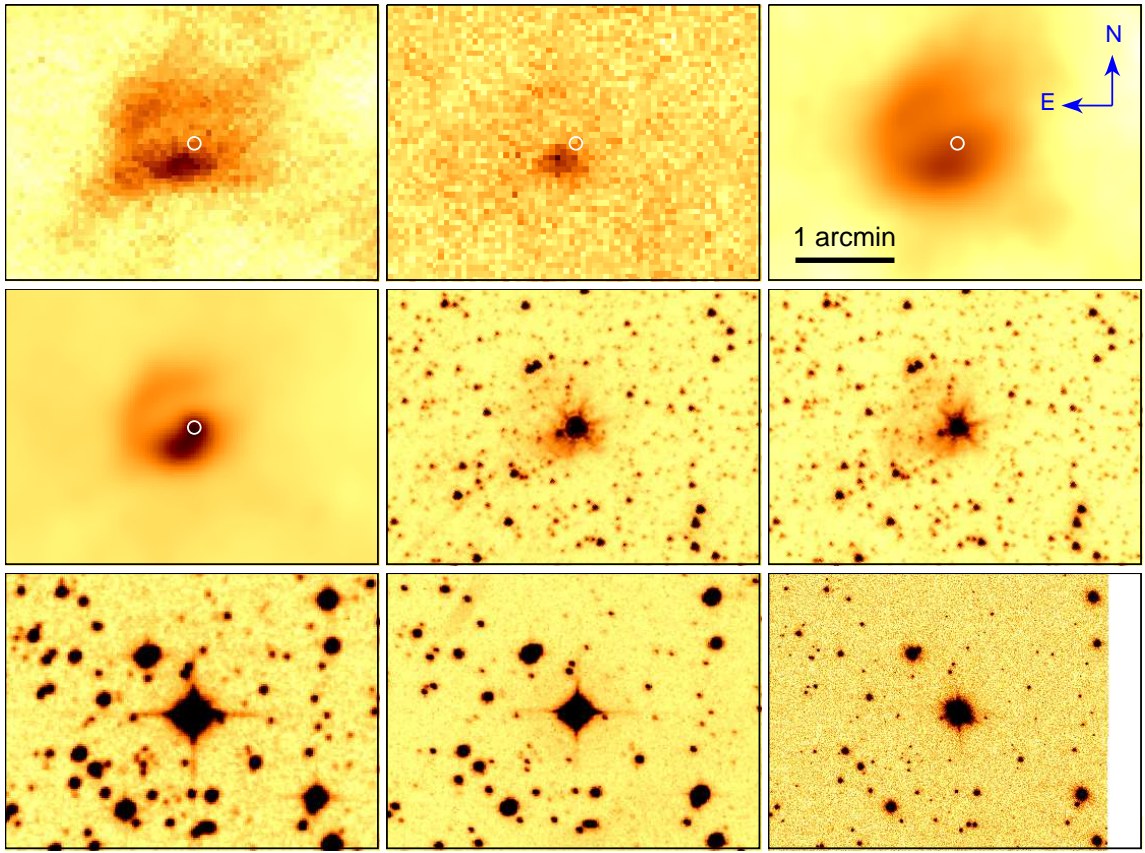}
\end{center}
\caption{From left to right and from top to bottom: {\it Herschel} 160 and 70\,\micron, {\it WISE} 22 and 12\,\micron, 
{\it Spitzer} 8 and 4.5\,\micron, DSS-II red-band, SHS H\,$\alpha$+[N\,{\sc ii}] and VPHAS+ H\,$\alpha$ images of 
TYC\,8606-2025-1 (marked by a circle) and its associated circumstellar nebula. A white stripe in the VPHAS+ image is 
a gap between the CCDs in the camera mosaic. A white stripe in the VPHAS+ image is a gap between the CCDs in the 
camera mosaic. The orientation and the scale of the images are the same. At the distance 
to TYC\,8606-2025-1 of 0.97 kpc, 1 arcmin corresponds to $\approx0.28$ pc.
    }
\label{fig:neb}
\end{figure*}

In the search for rare types of massive stars through the detection of their infrared (IR) circumstellar 
shells (e.g. Gvaramadze, Kniazev \& Fabrika 2010; Gvaramadze et al. 2012), we discovered numerous 
compact nebulae of various shapes using data from the {\it Spitzer Space Telescope} (Werner et al. 
2004) and the {\it Wide-field Infrared Survey Explorer} ({\it WISE}; Wright et al. 2010). 
Follow-up spectroscopy of their central stars showed that most of them are 
produced by luminous blue variables, blue supergiants, and Wolf-Rayet stars (e.g. Kniazev \& Gvaramadze 2015; 
Gvaramadze \& Kniazev 2017 and references therein). As a by-product, we found that several of the detected nebulae
were created in the course of evolution of low-mass stars (e.g. Gvaramadze \& Kniazev 2017; Gvaramadze et al. 
2019). One of these nebulae and its associated star TYC\,8606-2025-1 are the topic of this paper.

In Section\,\ref{sec:tyc}, we present multiwavelength images of the nebula and review the existing data on
TYC\,8606-2025-1. Spectroscopic observations and spectral analysis of TYC\,8606-2025-1 are described, respectively, 
in Sections\,\ref{sec:obs} and \ref{sec:spe}. In Section\,\ref{sec:dis}, we discuss the obtained results and 
outline further work. Finally, Section\,\ref{sec:sum} provides a summary.

\section{A spiral infrared nebula and its associated star TYC\,8606-2025-1}
\label{sec:tyc}

The nebula around TYC\,8606-2025-1 was discovered using data from the {\it WISE} mid-IR all sky survey 
(Wright et al. 2010). This survey provides images at four wavelengths: 22, 12, 4.6 and 3.4\,\micron, with 
an angular resolution of 6.1, 6.4, 6.5 and 12.0 arcsec, respectively. The nebula is clearly seen in the 22 
and 12\,\micron \, images (see Fig.\,\ref{fig:neb}), and at both wavelengths its angular size is 
$\approx1$ arcmin. The SIMBAD data base\footnote{http://simbad.harvard.edu/simbad/} indicates the 
{\it Infrared Astronomical Satellite} ({\it IRAS}) source IRAS\,09546$-$5634 at 16 arcsec from 
TYC\,8606-2025-1, but does not provide any bibliographic reference for it.  

The nebula attracted our attention because of its unusual appearance. Unlike other mid-IR nebulae 
discovered with {\it Spitzer} and {\it WISE}, most of which have circular or bipolar appearance 
(Gvaramadze et al. 2010), the nebula produced by TYC\,8606-2025-1 appears as a spiral. The spiral-like
shape of the nebula is more obvious in the {\it WISE} 12\,\micron \, image showing that the spiral starts 
from TYC\,8606-2025-1 and then bends around this star on the east side. The brightness distribution over 
the nebula is inhomogeneous, with its maximum close to the star and dropping by a factor of several at the 
periphery.

The nebula was also observed with the Infrared Array Camera (IRAC; Fazio et al. 2004) on board the {\it Spitzer Space 
Telescope} at 8 and 4.5\,\micron, and with the Photodetector Array Camera and Spectrometer (PACS) instrument on board 
the {\it Herschel} Space Observatory (Pilbratt et al. 2010) at 160 and 70\,\micron. Fig.\,\ref{fig:neb} shows that the nebula 
is visible (at least partially) at all these wavelengths. In the 160\,\micron \, image, it appears that the nebula is 
blended with back- or foreground emission. 

We also searched for an optical counterpart to the nebula using the Digitized Sky Survey\,II (DSS-II; McLean et al. 2000), 
the SuperCOSMOS H-alpha Survey (SHS; Parker et al. 2005) and the VST Photometric H\,$\alpha$ Survey of the Southern 
Galactic Plane and Bulge (VPHAS+; Drew et al. 2014), but did not find it (see, however, Section\,\ref{sec:neb}). 
Moreover, we did not find any signature of the nebula in the optical long-slit spectrum of TYC\,8606-2025-1 (see 
Section\,\ref{sec:obs}).

In Table\,\ref{tab:det}, we summarize some of the properties of TYC\,8606-2025-1. The spectral type is based on our 
spectroscopic observations and spectral analysis. The $B$ and $V$ magnitudes are from the Tycho-2 catalogue (H{\o}g et al. 2000).
The coordinates and the 
$JHK_{\rm s}$ photometry are taken from the Two-Micron All Sky Survey (2MASS; Skrutskie et al. 2006) All-Sky Catalog 
of Point Sources (Cutri et al. 2003). The {\it WISE} photometry is from the AllWISE Data Release (Cutri et al. 2014).
The distance is based on the {\it Gaia} second data release (DR2; Gaia Collaboration et al. 2018) parallax of 
TYC\,8606-2025-1 of $1.0340\pm0.0378$\,mas, which places this star at a distance of $d=0.97^{+0.04} _{-0.03}$ kpc. 
At this distance, the linear size of the nebula is $\approx0.28$ pc. 

\begin{table}
  \centering{\caption{Properties of TYC\,8606-2025-1.}
  \label{tab:det}
 \begin{tabular}{lc}
    \hline
  Spectral type & G8\,III Ba0 \\
  RA(J2000) & $09^{\rm h} 56^{\rm m} 23\fs02$  \\
  Dec.(J2000) & $-56\degr 48\arcmin 51\farcs0$ \\
  $l$ & $280\fdg6746$ \\
  $b$ & $-1\fdg7076$  \\
  $B$ (mag) & $11.42\pm0.07$ \\
  $V$ (mag) & $10.32\pm0.04$ \\
  $J$ (mag) & $8.330\pm0.027$ \\
  $H$ (mag) & $7.850\pm0.036$ \\
  $K_{\rm s}$ (mag) & $7.639\pm0.021$ \\
  $[3.4]$ (mag) & $7.512\pm0.028$ \\ 			
  $[4.6]$ (mag) & $7.648\pm0.020$ \\
  $[12]$ (mag) & $6.943\pm0.018$ \\
  $[22]$ (mag) & $5.852\pm0.044$ \\
  $d$ (kpc)    & $d=0.97^{+0.04} _{-0.03}$ \\
  \hline
 \end{tabular}
}
\end{table}

\section{Spectroscopic observations}
\label{sec:obs}

To get an idea on the nature of TYC\,8606-2025-1, we obtained its long-slit spectrum on 2014 April 11 with the Robert 
Stobie Spectrograph (RSS; Burgh et al. 2003; Kobulnicky et al. 2003) at the Southern African Large Telescope (SALT;  
Buckley, Swart \& Meiring 2006; O'Donoghue et al. 2006). The volume phase holographic (VPH) grating GR900 was used 
to cover the spectral range from 4200 to 7200~\AA, with a slit width of 1.25 arcsec and the spectral resolution of 
FWHM $\approx$4.5~\AA \, ($R\approx1000$). The slit was oriented at the position angle of 135\degr \, (measured from north 
to east), i.e. in such a way to cover the brightest part of the nebula. 

We obtained two short (30\,s) and two long (300\,s) exposures with an average seeing of 1.3 arcsec.
A reference spectrum of a Xe arc lamp was obtained immediately after each observation. The spectrophotometric standard 
star CD$-$32\degr9927 (Hamuy et al. 1994) was observed during twilight time for relative flux calibration (the 
absolute flux calibration is not possible with SALT because the unfilled entrance pupil of the telescope moves during 
the observations). 

The primary reduction of the RSS data was done with the SALT science pipeline (Crawford et al. 2010). The subsequent 
long-slit data reduction was carried out in the way described in Kniazev et al. (2008). 

To search for possible radial velocity variability and for spectral modelling, we obtained five  
spectra of TYC\,8606-2025-1 with the SALT High Resolution Spectrograph (HRS; Barnes et al. 2008; Bramall et al. 
2010, 2012; Crause et al. 2014) in the medium resolution (MR) mode on 2018 December 20, 22 and 25, 2019 January 18 
and April 17 with an exposure time of each observation of 400\,s. The seeing in these observations was 2.2, 1.6, 
1.5, 1.7 and 1.8 arcsec, respectively. The signal-to-noise ratio (SNR) in these spectra varied between $\approx100$
and 150.

The HRS is a dual beam, fibre-fed \'echelle spectrograph. In the MR mode it has 2.23 arcsec diameter for both the 
object and sky fibres providing a spectrum in the blue and red arms over the spectral range of $\approx$3700--8900~\AA\ 
with resolving power of $R\approx36\,500-39\,000$ (Kniazev et al. 2019). In our observations, both the blue and red arm 
CCDs were read out by a single amplifier with a 1$\times$1 binning. A spectrum of a ThAr lamp and three spectral 
flats were obtained in the MR mode during a weekly set of HRS calibrations.

The primary reduction of the obtained \'echelle spectra was performed with the SALT science pipeline (Crawford et al. 2010). 
The subsequent reduction steps, which include background subtraction, spectral order extraction, removal of the blaze 
function, identification of the arc lines, wavelength calibration, and merging of spectral orders, were carried out using 
the MIDAS HRS pipeline described in detail in Kniazev, Gvaramadze \& Berdnikov (2016) and Kniazev et al. (2019).

We also retrieved processed spectra of TYC\,8606-2025-1 from the European Southern Observatory (ESO) data archive. 
They were obtained on 2015 May 5 under programme ID 095.A-9011(A) with the Fibre-fed Extended Range Optical Spectrograph 
(FEROS; Kaufer et al. 1999) --- a high-resolution ($R=48\,000$) \'echelle spectrograph --- mounted on the 2.2-m Max 
Planck Gesellschaft telescope at La Silla. Four spectra covering the spectral range of $\approx3500-9200$~\AA \, were 
taken with a total exposure time of 6800~s. The spectra were reduced with the ESO pipeline for Phase~3, co-added and 
normalized to the continuum. The SNR of the resulting spectrum is $\approx150$.

\section{Spectral analysis}
\label{sec:spe}

\subsection{RSS spectrum}
\label{sec:rss}

The RSS spectrum of TYC\,8606-2025-1 does not show emission lines and is typical
of cool stars of the G--K spectral type. The H$\gamma$/Fe\,{\sc i} $\lambda$4325 intensity ratio of $<1$
in this spectrum implies that TYC\,8606-2025-1 is of G8 spectral type (Evans et al. 2004).  

No nebular emission lines were detected in the 2D RSS spectrum, neither at the position of the IR nebula, nor 
along the whole 8 arcmin long spectroscopic slit.

\subsection{HRS spectra}
\label{sec:hrs}

\subsubsection{Atmospheric parameters and abundances}
\label{abu}

We used the SME (Spectroscopy Made Easy) package (Valenti \& Piskunov 1996; Piskunov \& Valenti 2017) for the atmospheric 
parameter determination and abundance estimates. This package was designed for the analysis of stellar spectra 
using spectral fitting techniques. After correction for heliocentric velocity and radial velocity variations (see 
Section\,\ref{sec:rad}), the HRS spectra were co-added to increase the SNR. 
Four spectral regions 5100--5250, 5600--5705, 6100--6250, and 6450--6630 \AA\, were 
chosen for the fitting. This choice was based on the presence of spectral lines sensitive to different atmospheric 
parameters (Mg\,{\sc i} triplet at 5167-5183~\AA, H$\alpha$ line wings, etc), as well as on the presence of 
molecular lines of C$_2$ and CN used for carbon and nitrogen abundance estimates. 

The fundamental atmospheric parameters, effective temperature $T_{\rm eff}$, surface gravity $\log g$, metallicity [Fe/H], 
micro- and macroturbulent velocities $V_{\rm mic}$ and $V_{\rm mac}$, and projected rotational velocity $v\sin i$, were 
derived during the first step of the SME analysis by varying the above-mentioned parameters. As a first approximation, 
we used the line masks defined in Ryabchikova et al. (2016). These masks were further corrected for defects (e.g., cosmics) 
in the observed spectrum of TYC\,8606-2025-1. The model atmosphere parameters were searched within a grid of the MARCS 
spherical atmosphere models (Gustafsson et al. 2008). The derived parameters together with the error estimates are given 
in Table~\ref{tab:param}. The error estimates are based on fit residuals, partial derivatives, and data uncertainties, as 
described in detail in Ryabchikova et al. (2016) and Piskunov \& Valenti (2017).

\begin{table}
	\centering\caption{Atmospheric parameters of TYC\,8606-2025-1 based on the HRS spectra.}
	\begin{tabular}{lc}
		\hline
		Parameter  & Value \\
		\hline
		$T_{\rm eff}$ (K)       & $4900\pm30$   \\                     
		$\log g$                & $2.31\pm0.16$ \\                      
        $[$Fe/H$]$              & $-0.10\pm0.04$ \\                      
		$v\sin i \, (\kms)$     & $0.64\pm3.78$  \\                      
		$V_{\rm mic} \, (\kms)$ & $1.43\pm0.10$ \\                       
		$V_{\rm mac} \, (\kms)$ & $3.71\pm1.11$ \\                     
		\hline
	\end{tabular}
	\label{tab:param}
\end{table}

In the second step of the SME analysis, we fixed all parameters derived during the first step and varied the abundances 
of the individual elements using the same masks. For a few elements we extracted the lines of the particular element from 
the common masks and ran SME with these individual masks. The abundances of O, Sr, La, and Eu were derived from the spectral 
synthesis of individual lines not falling into the chosen SME spectral regions: the O\,{\sc i} IR triplet 
$\lambda\lambda$7771.94, 7774.16 and 7775.39 \AA, Sr\,{\sc i} $\lambda\lambda$4607.33 and 7070.07 \AA, La\,{\sc ii} 
$\lambda\lambda$5114.56, 5301.97, 5303.55, 5805.78 and 5808.32 \AA, and Eu\,{\sc ii} $\lambda\lambda$6437.64 and 6645.11 
\AA. The carbon and nitrogen abundances were estimated by simultaneous fitting of the C$_2$ and CN molecular lines distributed 
in all regions covered by the SME analysis.     

\begin{table}
	\begin{footnotesize}
	\centering\caption{Surface element (X) abundances of TYC\,8606-2025-1 and the Sun in $\log$(X/H)+12 units (second and 
	third columns, respectively). The fourth column lists abundances relative to the Sun, [X/H]=log(X/H)$-$log(X$_\odot$/H$_\odot$). 
	The last column provides information about different effects taken into account in deriving abundances.}
		\label{tab:abun}
		\begin{tabular}{lccrc}
			\hline
			 \multicolumn{1}{l}{X} & \multicolumn{1}{c}{Star} & \multicolumn{1}{c}{Sun} & \multicolumn{1}{c}{[X/H]} & Remark  \\                      
			\hline
			C     & $8.11 \pm0.04$ & $8.43 $  & $-0.32$ & C$_2$ \\                                     
			N     & $8.37 \pm0.06$ & $7.83 $  & 0.54 & CN \\                                         
			O**   & $8.71 \pm0.09$ & $8.69 $  & 0.02 & NLTE\\ 
			Na*   & $6.53 \pm0.06$ & $6.21 $  & 0.32 & NLTE, hfs \\  
			Al    & $6.21 \pm0.02$ & $6.37 $  & $-0.16$ & \\  
			Mg    & $7.49 \pm0.10$ & $7.59 $  & $-0.10$ &  \\                                           
			Si*   & $7.52 \pm0.07$ & $7.51 $  & 0.01 &  \\                                           
			Ca*   & $6.39 \pm0.05$ & $6.32 $  & 0.07 &  \\                                           
			Sc*   & $3.03 \pm0.04$ & $3.16 $  & $-0.13$ & hfs \\                                        
			Ti    & $4.77 \pm0.06$ & $4.93 $  & $-0.16$ &    \\                                         
			V*    & $3.79 \pm0.04$ & $3.89 $  & $-0.10$ & hfs \\                                        
			Cr    & $5.47 \pm0.06$ & $5.62 $  & $-0.15$ &  \\                                           
			Mn    & $5.19 \pm0.08$ & $5.42 $  & $-0.23$ & hfs \\                                        
			Fe    & $7.37 \pm0.07$ & $7.47 $  & $-0.10$ &  \\                                           
			Co*   & $4.69 \pm0.02$ & $4.93 $  & $-0.24$ & hfs \\                                        
			Ni    & $6.06 \pm0.06$ & $6.20 $  & $-0.14$ &  \\                                           
			Sr**  & $3.14 \pm0.21$ & $2.83$  & 0.31 & NLTE \\ 
			Y     & $2.41 \pm0.08$ & $2.21 $  & 0.20 & hfs \\                                        
			Zr*   & $2.77 \pm0.04$ & $2.59 $  & 0.18 &  \\                                           
			Ba    & $2.80 \pm0.06$ & $2.25 $  & 0.55 & hfs, IS \\                                    
			La**  & $1.33 \pm0.03$ & $1.11 $ & 0.22 &  \\                                           
			Ce*   & $1.83 \pm0.05$ & $1.58 $  & 0.25 &  \\                                           
			Nd*   & $1.54 \pm0.06$ & $1.42 $  & 0.12 &  \\                                           
			Eu**  & $0.54 \pm0.13$ & $0.52 $  & 0.02 & hfs, IS \\ 
			\hline                                     
 \multicolumn{5}{p{.4\textwidth}}{{\it Note}: One asterisk marks the elements for which abundances were derived 
 using individual line masks in SME. Two asterisks mark the elements whose abundances were derived by spectral 
 synthesis of individual lines. In the latter case, the uncertainty corresponds to the standard deviation of the 
 single line abundances.} \\		
    \end{tabular}
	\end{footnotesize}
	\end{table}
 
The atomic parameters of the lines were extracted from the Vienna Atomic Line Database ({\sc vald}; Kupka et al. 1999)
using its 3rd release ({\sc vald3}; Ryabchikova et al. 2015). The atomic parameters for the C$_2$ and CN molecular lines 
were taken from Brooke et al. (2013) and Kurucz (2010), respectively.
The isotopic structure (IS) of Cu\,{\sc ii}, Ba\,{\sc ii}, and Eu\,{\sc ii} was taken into account. For elements with  
odd isotopes the synthetic calculations included hyperfine splitting (hfs). Most of the data for isotopic and 
hyperfine splitting were obtained from the website of R.~L.~Kurucz\footnote{http://kurucz.harvard.edu/atoms.html}, 
who collected data and a bibliography for individual atomic species. In particular, the hyperfine structure constants for 
Na\,{\sc i} were taken from Griffith et al. (1977), Tsekeris et al. (1976), and Yei et al. (1993), for Sc\,{\sc i} and
{\sc ii} from Ertmer \& Hofer (1976) and Mansour et al. (1989), for V\,{\sc i} from Childs et al. (1979), Palmeri et al. 
(1995), and Unkel et al. (1989), for Mn\,{\sc i}, {\sc ii} from Blackwell-Whitehead et al. (2005) and Holt et al. (1999), 
for Co\,{\sc i} from Pickering (1996), for Cu\,{\sc i} from Fischer et al. (1967) and Ney (1966), for Y\,{\sc i} from 
Dinneen et al. (1991) and Villemoes et al. (1992), and for Ba\,{\sc i} from Villemoes et al. (1993). 
All hfs data were collected in a new database (Pakhomov et al., in preparation) and used by the internal {\sc vald} 
programs for the hfs calculations.

\begin{figure}
	\includegraphics[width=\columnwidth,clip=]{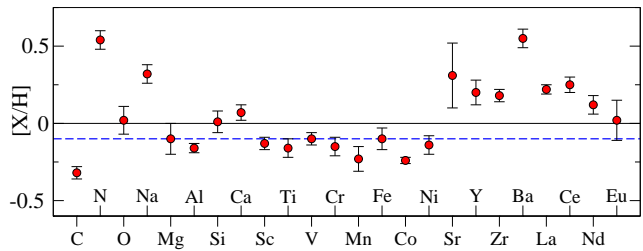}
	\caption{Elemental abundances of TYC\,8606-2025-1 relative to the Sun. The (black) solid line corresponds to 
	solar abundance values. The metallicity [Fe/H] level is marked by a (blue) dashed line. 
	}
	\label{fig:abund}
\end{figure}

\begin{figure*}
	\includegraphics[width=12cm]{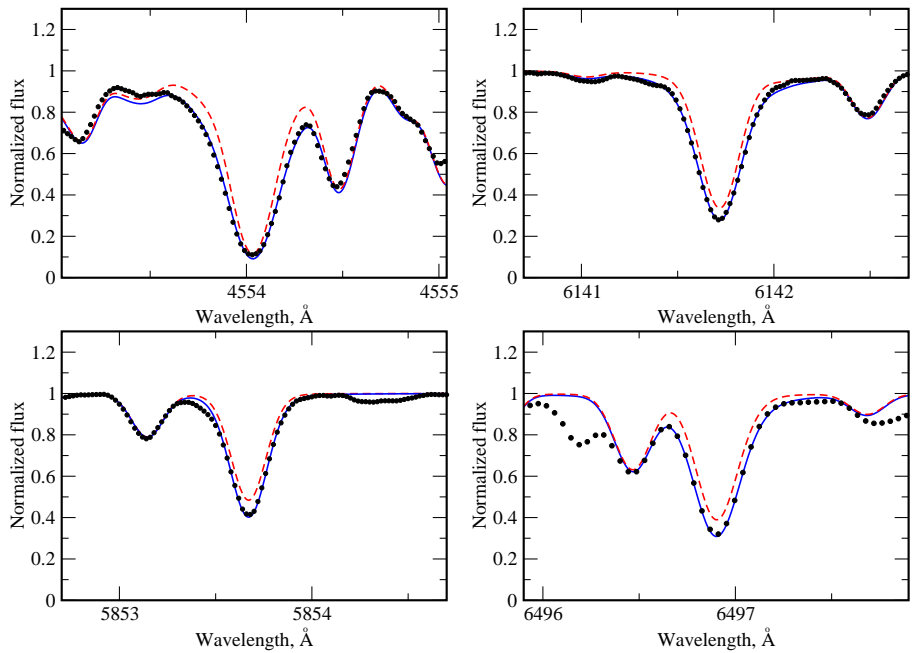}
	\caption{Portions of the HRS spectrum of TYC\,8606-2025-1 with Ba\,{\sc ii} lines used to determine
	the abundance of this element. The observed spectra (black dots) are compared with synthetic NLTE spectra 
	calculated from the derived atmospheric parameters for the solar abundances (dashed red line) and the
	final estimated abundances (solid blue line). One can see that the solar Ba abundance is not 
	a good match to the observed spectrum, indicating that Ba is overabundant.
	}
	\label{fig:BaII}
\end{figure*}
%
\begin{figure}
	\includegraphics[width=\columnwidth,clip=]{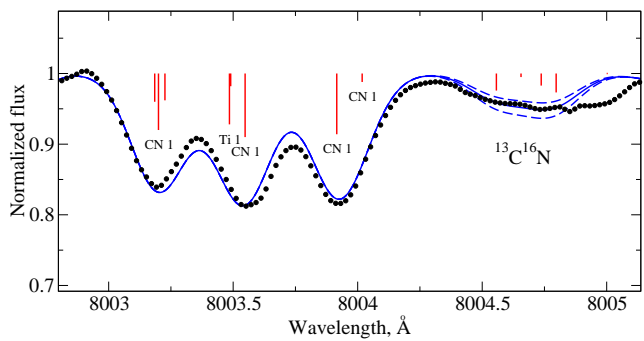}
	\caption{Profiles of the CN molecular lines in the HRS spectrum of TYC\,8606-2025-1 (black dots). The observed
	 profiles (black dots) are 	compared with models (blue lines), of which the solid one corresponds to 
	 $^{12}$C/$^{13}$C=14, while the dashed ones correspond to this ratio changed by $\pm$4.}
	\label{fig:13C}
\end{figure}

The average abundances for the final atmospheric model of TYC\,8606-2025-1 are given in $\log$(X/H)+12 units in the second
column of Table~\ref{tab:abun}. The third column of this table presents the current solar photospheric abundances (Asplund 
et al. 2009; Scott et al. 2015a,b; Grevesse et al. 2015), while the fourth columns lists the elemental abundances derived 
for TYC\,8606-2025-1 relative to the Sun. The last column of Table~\ref{tab:abun} provides information about different 
effects taken into account in the SME or spectral synthesis analyses. For clarity, the elemental abundances of 
TYC\,8606-2025-1 relative to the Sun are shown in Fig.~\ref{fig:abund}, highlighting the overabundance of s-process 
elements, with the Ba abundance increased by a factor of about three with respect to solar. 

The current analysis was performed in the LTE approximation. In principle, SME allows one to include NLTE calculations in 
the spectral fitting (Piskunov et al. 2017), but this requires to provide NLTE departure coefficients for the grid 
calculations. At present, this is done for the MARCS grid of plane-parallel models. However, for a few elements NLTE 
corrections were applied to the LTE abundances. An average NLTE correction for the IR O\,{\sc i} triplet of $-0.27$~dex 
was calculated for us by T.M.\,Sitnova following a model atom from Sitnova \& Mashonkina (2018). For Na\,{\sc i}, a NLTE 
correction of $-0.1$~dex was estimated from the work by Alexeeva et al. (2014). A NLTE correction of +0.18~dex was taken 
from Bergemann et al. (2012) and applied to the LTE abundance of Sr of 2.81~dex based on the Sr\,{\sc i} 4607.331~\AA\, 
line. The corrected abundance of 2.99~dex agrees with the LTE abundance of 3.29 dex derived from the weaker Sr\,{\sc i} 
line at 7070.071 \AA\, for which no NLTE correction was estimated by Bergemann et al. (2012). The resulting Sr abundance 
given in Table\,\ref{tab:abun} is the mean of the two estimates. The NLTE correction for Ba does not exceed $-0.1$~dex 
(L.I.\,Mashonkina, private communication). Fig.~\ref{fig:BaII} shows all Ba lines used to derive the abundance of this 
element. One can see that all of them clearly indicate that Ba is overabundant.

We checked for the possible presence of technetium lines in the spectrum of TYC\,8606-2025-1. This element is observed 
in spectra of (single) AGB stars, which are luminous enough to self-enrich themselves with heavy elements (e.g. Lebzelter 
\& Hron 2003). We searched for the Tc\,{\sc i} resonance lines at 4238.19, 4262.27, and 4297.06~\AA. All of them are strongly 
blended with other absorption lines, which could successfully be fitted without adding the Tc\,{\sc i} lines, indicating that 
there is no sign of the presence of this element in the spectrum.
The non-detection of technetium implies  
that the enhanced abundances of the s-process elements in TYC\,8606-2025-1 are likely due to mass transfer from a (more 
evolved) companion star (see Section\,\ref{sec:bar}).

We also estimated the ratio $^{12}$C/$^{13}$C, which is an indicator of the stellar evolutionary status (Iben 1966). 
For this we used the spectral region around $\lambda$8000~\AA, which is devoid of lines except of $^{13}$C$^{16}$N. 
We used the molecular data from Sneden et al. (2014) to create a list of $^{13}$C$^{16}$N lines. The observed spectrum 
was fitted with a synthetic one as shown in Fig.~\ref{fig:13C}. We found that $^{12}$C/$^{13}$C=$14\pm4$, implying
that TYC\,8606-2025-1 is a giant star (e.g. Dearborn, Eggleton \& Schramm 1976). 

\subsubsection{Radial velocity measurements for TYC\,8606-2025-1}
\label{sec:rad}

We measured the heliocentric radial velocity, $V_{\rm r,hel}$, of TYC\,8606-2025-1 from each of the five available 
HRS spectra. The obtained results are given in Table\,\ref{tab:rad}, to which we also added the radial velocity derived 
from the FEROS spectrum. From Table\,\ref{tab:rad} it follows that $V_{\rm r,hel}$ has changed by about $2 \, \kms$ 
during the last four years, suggesting that TYC\,8606-2025-1 is a binary system. Note also that the positive sign of 
$V_{\rm r,hel}$ in 2015 is consistent with that of the {\it Gaia} DR2 median (barycentric) radial velocity of 
TYC\,8606-2025-1 of $+1.30\pm0.20 \, \kms$, based on six measurements carried out during a period of 22 months 
between 25 July 2014 and 23 May 2016. Unfortunately, the individual {\it Gaia} measurements of the radial velocity 
are not available in DR2 (Sartoretti et al. 2018), which did not allow us to attempt constraining the orbital parameters 
of TYC\,8606-2025-1 (see, however, Section\,\ref{sec:neb}).

\begin{table}
\centering\caption{Changes in the heliocentric radial velocity of TYC\,8606-2025-1.}
\label{tab:rad}
\begin{tabular}{lcc} \hline
Date & JD & $V_{\rm r,hel} (\kms)$ \\
\hline
2015 May 6       & 2457148.0 & $+0.6\pm0.1$ \\
2018 December 20 & 2458473.5 & $-0.2\pm0.2$ \\
2018 December 22 & 2458475.5 & $-0.2\pm0.2$ \\
2018 December 25 & 2458478.5 & $-0.4\pm0.2$ \\
2019 January 18  & 2458502.4 & $-0.6\pm0.2$ \\
2019 April 17    & 2458591.4 & $-1.6\pm0.2$ \\
\hline
\end{tabular}
\end{table}

\subsection{Reddening and luminosity of TYC\,8606-2025-1}

\begin{figure}
	\includegraphics[width=\columnwidth,clip=]{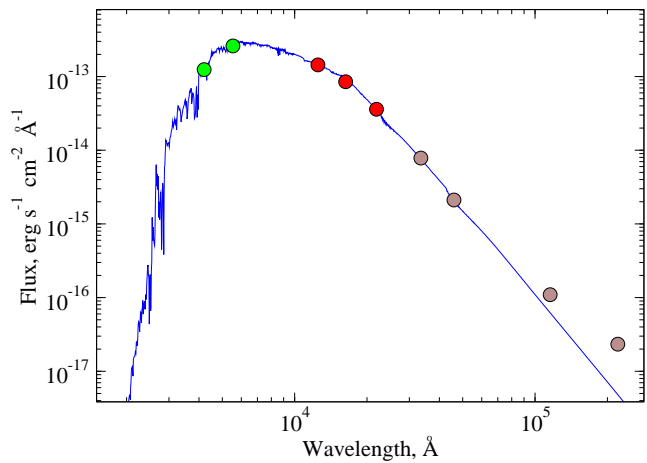}
	\caption{Observed flux distribution of TYC\,8606-2025-1 (filled colour circles) based on the photometric data 
	compiled in Table\,\ref{tab:det}, compared to the model continuum (blue line).
	}
	\label{fig:sed}
\end{figure}

Using photometric data from Table\,\ref{tab:det}, we fit the observed spectral energy distribution (SED) of 
TYC\,8606-2025-1 with a model SED (see Fig.~\ref{fig:sed}), calculated by interpolation of grids of ATLAS9 model 
atmospheres\footnote{http://wwwuser.oats.inaf.it/castelli/grids.html} by Castelli \& Kurucz (2003) and scaled to 
$d=0.97$ kpc according to the parameters given in Table\,\ref{tab:param}. We adopted the extinction law of Mathis 
(1990) with a total-to-selective absorption ratio of $R_V=3.1$. The best fit was achieved with the colour excess
$E(B-V)=0.14\pm0.01$\,mag. The excess flux seen at 12 and 22 \micron \, is obviously due to nebular emission, since 
the point spread function of the {\it WISE} instrument at these wavelengths of $\approx6$ and 12 arcsec, respectively, 
covers a part of the nebula.

\begin{figure}
	\includegraphics[width=\columnwidth,clip=]{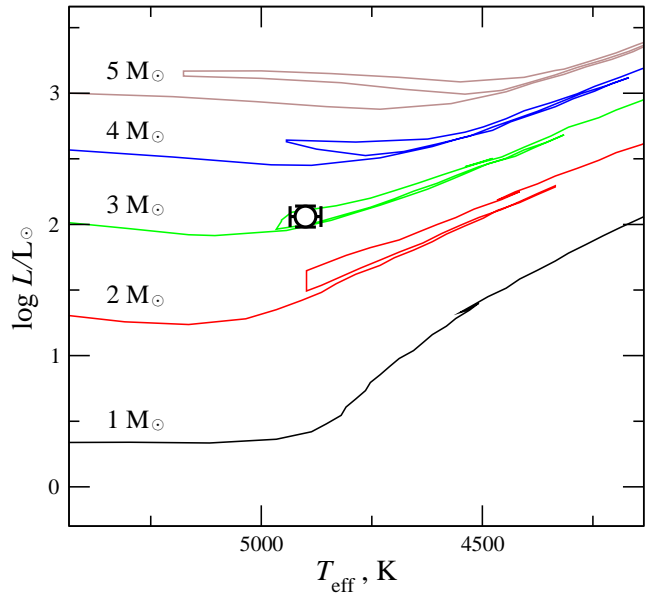}
	\caption{Position of TYC\,8606-2025-1 in the Hertzsprung-Russell diagram (circle with error bars). 
	Also shown are the evolutionary tracks of Girardi et al. (2000). 
	}
	\label{fig:HR}
\end{figure}

Then, using the bolometric correction of $-0.28$\,mag from Bessell, Castelli \& Plez (1998), we derive the absolute 
visual magnitude and luminosity of the star of $M_V \approx -0.04\pm0.09$\,mag and $\log(L/\lsun)\approx 2.02\pm0.04$, 
respectively, which correspond to an initial mass of the star of $\approx 3 \, \msun$ (e.g. Girardi et al. 2000; 
Ekstr\"om et al. 2012; see also Fig.~\ref{fig:HR}). 

\section{Discussion}
\label{sec:dis}

\subsection{TYC\,8606-2025-1 as a mild barium star}
\label{sec:bar}

Inspection of Table\,\ref{tab:abun} shows (see also Fig.\,\ref{fig:abund}) that the surface carbon and nitrogen
abundances in TYC\,8606-2025-1 are, respectively, depleted and enhanced by factors of two and three, while the 
oxygen abundance is equal to the solar one within 1$\sigma$. These abundances are typical of normal G and
K giants and could be understood as the result of first dredge-up occurring on the red-giant branch (Lambert \& Ries 1981). 
Indeed, according to the evolutionary tracks by Ekstr\"om et al. (2012), a $3\,\msun$ star reaches $T_{\rm eff}=4900$\,K the 
first time evolving redward, after core hydrogen exhaustion. At this time the surface abundances remain unchanged. 
Then, during the core helium-burning, the star undergoes a blue loop and goes back to $T_{\rm eff}=4900$\,K. 
At this stage, the CN-processed material is dredged-up to the surface of the star, with the C and N abundances,
correspondingly, reduced and increased by factors of 1.7 and 3, and O remaining unchanged.

Table\,\ref{tab:abun} and Fig.\,\ref{fig:abund} also show an excess of elements produced in the slow neutron-capture
nucleosynthesis (s-process elements), such as strontium, yttrium, zirconium, barium, lanthanum and cerium, of which the most 
overabundant (by a factor of $\approx3$) is Ba. The enhanced Ba abundance is typical of the so-called barium stars 
(first recognized as a spectroscopic class by Bidelman \& Keenan 1951). These (classical) barium stars, however, exhibit 
an increased carbon abundance and much stronger excess of the s-process elements with Ba enhanced by up to a factor of 10--20. 

Observations show that all classical barium stars are binary systems (McClure, Fletcher \& Nemec 1980; McClure \& Woodsworth 
1990; Jorissen et al. 2019), meaning that binarity is a prerequisite for their formation. It is generally accepted 
that barium stars are post mass-transfer binaries composed of a giant (or dwarf) G--K star and a white dwarf (McClure et 
al. 1980; Boffin \& Jorissen 1988). Detection of white dwarf companions to some barium stars lends weighty support to this 
view (B\"ohm-Vitense 1980; Schindler et al. 1982; Gray et al. 2011). Barium and other s-process elements are synthesized 
during the thermally pulsing asymptotic giant branch (TP-AGB) evolutionary phase of the primary star (now white dwarf) and 
are transferred by stellar wind to the companion star, thereby polluting its surface with heavy elements. In most cases, 
however, the primary star has faded in the visual and IR wavelengths below detectability, and its presence can only be revealed by 
UV (B\"ohm-Vitense 1980; Schindler et al. 1982) and/or X-ray observations (Schindler et al. 1982; Jorissen et al. 1996), or 
through indirect signs, such as a radial velocity variability in the companion (barium) star (e.g. McClure et al. 1980) or 
detection of an optically visible planetary nebula (PN) around a cool chemically peculiar star (e.g. Miszalski et al. 2013).

The moderate enhancement of barium and other s-process elements and the depleted carbon abundance in TYC\,8606-2025-1 
imply that this star is a mild (or marginal) barium star (MacConnell, Frye \& Upgren 1972; Pilachowski 1977; Boyarchuk 
et al. 2002). Following the spectral classification of Morgan \& Keenan (1973), we assign to TYC\,8606-2025-1 
the barium index (a characteristic of barium enhancement) of Ba0, so that it is a G8\,III Ba0 star. Although the origin of 
mild barium stars is still open to debate, it is possible that the relatively low excess of s-process elements in these 
stars is at least partly a consequence of a larger separation between companion stars, reducing the efficiency of wind 
accretion (B\"ohm-Vitense, Nemec \& Proffitt 1984). This possibility is consistent with the observed trend in barium 
stars towards a decrease of the s-process element excess with the increase of the binary period (see table\,8 in Jorissen 
et al. 2019). Interestingly, stars with large barium indices form a population of binary systems with nearly circular, 
short period ($<1000$\,d) orbits, while those with lower barium indices mostly reside in eccentric binaries with longer 
(up to a few 10\,000\,d or $\sim100$\,yr) orbital periods (Jorissen et al. 2019). The combination of large orbital 
periods and eccentricities is still unclear.

The detection of radial velocity variations in TYC\,8606-2025-1 (indicating that it is a binary system) and the luminosity 
of this star (not high enough for any dredge up of the s-process elements from the stellar interior to occur) suggest that, 
like in the classical barium stars, the surface of TYC\,8606-2025-1 was polluted with heavy elements via wind mass transfer 
from the companion star during its TP-AGB phase. Moreover, from the fact that TYC\,8606-2025-1 is in the core He-burning 
stage and because the post-AGB phase is very short ($\sim10\,000$\,yr), it is likely that the initial mass ratio of the 
binary system was close to unity (within about 10 per cent), meaning that the initial mass of the primary star was $\approx3 
\, \msun$. 
 
Currently only few classical and mild barium stars are known to be associated with PNe (Th\'evenin \& Jasniewicz 1997; 
Bond, Pollacco \& Webbink 2003; Miszalski et al. 2012, 2013; Tyndall et al. 2013; L\"obling, Boffin \& Jones2019), 
and for some of them a hot companion star (white dwarf) has been detected in the UV. Moreover, for one of these 
stars the orbital period of $\approx2700$\,d was measured (Jones et al. 2017; Aller et al. 2018). Unlike the nebula around 
TYC\,8606-2025-1, all known nebulae around barium stars are visible both at optical 
and infrared wavelengths. Optical spectroscopy of several of them shows rich emission spectra dominated 
by Balmer and helium lines, and very strong forbidden lines of oxygen and nitrogen (Miszalski et al. 2012, 2013).

The non-detection of an optical counterpart to the IR nebula around TYC\,8606-2025-1 could indicate that the primary star 
in this system has only recently left the AGB phase and is not hot enough ($T_{\rm eff}<30\,000$\, K) to ionize its 
surroundings. In this case, however, the primary star would be visually 2--4 mag brighter than TYC\,8606-2025-1 because  
on the horizontal part of the post-AGB track its luminosity of $\log(L/\lsun)\approx4$ (e.g. Bl\"ocker \& Sch\"onberner 
1997) would be two orders of magnitude higher than that of the G star. It is therefore more likely that the primary star 
has already exposed its hot ($\sim100\,000$\,K) core of $\approx0.6 \, \msun$ (e.g. Bl\"ocker \& Sch\"onberner 1997) and 
that it is currently on the white dwarf cooling track. In this case, the star would contribute mostly to the UV part of 
the SED of TYC\,8606-2025-1, while at optical wavelengths it would be several magnitudes fainter than the G star. 
This, however, raises a question: Why does not the spiral nebula have an optical counterpart and did not exhibit evidence of 
nebular emissions in the 2D RSS spectrum in spite of the low reddening of its associated star? 

\subsection{Nebula around TYC\,8606-2025-1}
\label{sec:neb}

The inference that TYC\,8606-2025-1 is a binary system composed of a G8 giant star and a hot primary star (white dwarf) 
implies that the origin of the spiral nebula should be somehow related to the mass lost from the latter star either during 
its AGB or post-AGB evolution. 

If the nebula was formed during the AGB and/or early post-AGB phase, then it should shine at optical wavelengths as a PN
because the wind material ejected at these phases is hydrogen-rich. The lack of optical emission suggests that the nebula 
is composed of hydrogen-poor material and that the oxygen, nitrogen and neon atoms (responsible for the strongest optical 
forbidden emission lines in spectra of PNe) are at least triply ionized and therefore did not form nebular lines at optical 
wavelengths (e.g. Gurzadyan 1970). This would imply that the observed IR emission from the nebula in the {\it WISE} 22 
and 12\,\micron \, bands is due to [O\,{\sc iv}] 25.9\,\micron \, and [Ne\,{\sc v}] 14.3 and 24.3\,\micron \, line 
emission, while at longer wavelengths it is due to dust emission (Chu et al. 2009; Flagey et al. 2011). Also, since the nebula 
is visible (in part) in both 8 and 4.5\,\micron \, IRAC bands, and the 4.5\,\micron \, band does not include PAH 
(polycyclic aromatic hydrocarbon) features, it is likely that at these wavelengths the nebular emission is also due to 
forbidden lines from highly ionized species (Chu et al. 2009).

For this explanation to work, the nebula should have formed after the primary star has experienced a very late thermal 
pulse (VLTP), i.e. a thermal pulse that occurred when the star was on the white dwarf cooling track and hydrogen 
burning was already off. In this case, the pulse-driven convection can led to a mixing and total burning of the 
remaining hydrogen shell in the star (Fujimoto 1977; Sch\"onberner 1979; Iben 1984; Herwig et al. 1999). As a result,
the star expands and cools, and rapidly returns to the AGB. During this ``born-again" episode the star develops a 
hydrogen-poor stellar wind, which could produce a new circumstellar nebula embedded in the more extended
hydrogen-rich material shed during the AGB phase (e.g. Wesson et al. 2008). Such nebulae were detected around several 
hydrogen-poor central stars of PNe (e.g. Zijlstra 2002 and references therein).

The post-VLTP evolution is very fast and 
proceeds on a time scale of $\sim100$\,yr. After the brief AGB phase the star undergoes a final blue loop and finds 
itself again on the white dwarf cooling track as a very hot bare stellar nucleus. The UV photons emitted by this
nucleus are hard enough to triply ionize metals in the newly formed hydrogen-poor nebula. 

The shape of the IR nebula around TYC\,8606-2025-1 also suggests that this star has recently gone through a brief 
episode of mass loss. Like in the case of spiral circumstellar nebulae produced by AGB stars (e.g. Mauron \& Huggins 
2006; Dinh-V.-Trung \& Lim 2009), it could be due to the orbital motion of the mass-losing star (Soker 1994; Mastrodemos 
\& Morris 1999). The stellar wind from such a star orbiting a windless companion star interacts with itself and induces 
a one arm spiral shock (Cant\'o et al. 1999; Wilkin \& Hausner 2017), which for circular orbits appears as an 
Archimedean spiral if the line of sight crosses the orbital plane at the right angle. The distance between successive 
turns of the spiral depends on the binary orbital period ($P$), and the orbital ($v_{\rm orb}$) and wind ($v_\infty$) 
velocities (Kim \& Taam 2002; Wilkin \& Hausner 2017). In the case of a fast wind ($v_\infty >>v_{\rm orb}$) this 
separation is simply equal to the product of the stellar wind velocity and the orbital period (Cant\'o et al. 1999; 
Wilkin \& Hausner 2017).

The IR nebula associated with TYC\,8606-2025-1 appears as a spiral that can be traced only about one turn around the 
star. This indicates that the nebula was formed during one orbital period. Since the G companion star could safely be 
considered windless and assuming that during the blue loop phase the wind velocity of the post-VLTP star was 
$v_\infty\sim1000 \, \kms$, one finds that in order to explain the separation of the spiral arm from the star of 
$\sim0.1$\,pc the orbital period of the binary should be $\sim100$\,yr. This value is at the upper end of the range of 
orbital periods derived for barium stars (Jorissen et al. 2019), and of the same order of magnitude as the time it takes 
the star after a VLTP to return in the white dwarf cooling track. 

Although the existing data did not allow us to derive the orbital parameters of TYC\,8606-2025-1, we note that 
these data are in agreement with our suggestion of a binary orbital period as long as $\sim100$\,yr. The spiral shape 
of the nebula implies that the orbital plane of the binary system makes a moderate angle, $\theta$, with the plane of 
the sky. Assuming that $\theta =30\degr$ and $P=100$\,yr, one finds that the observed values of $V_{\rm r,hel}$ could 
be fit with an eccentric orbit with an eccentricity of $>0.7$, which agrees with the observed tendency in barium stars to 
have more eccentric orbits with the increase of the orbital period (Jorissen et al. 2019; see their fig.\,7).

\begin{figure}
	\includegraphics[width=\columnwidth]{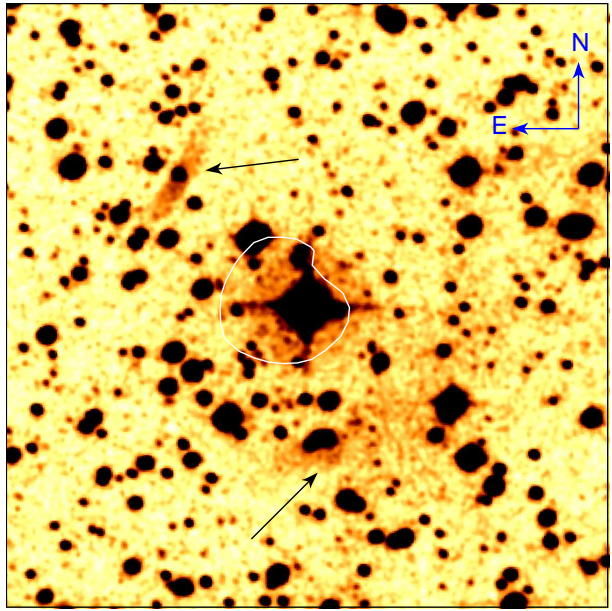}
	\caption{SHS H\,$\alpha$+[N\,{\sc ii}] image of the 5 arcmin $\times$ 5 arcmin field centred on TYC\,8606-2025-1.
	The optical features possibly associated with the star are indicated by arrows. The position of the IR 
	nebula is shown by a white contour. See text for details. At the distance to TYC\,8606-2025-1 of 0.97 kpc, 
	1 arcmin corresponds to $\approx0.28$ pc.
	}
	\label{fig:fil}
\end{figure}

If our suggestions on the origin of the IR nebula are correct, then one can expect that the hydrogen-rich material 
ejected by the primary star during the AGB and early post-AGB phases should be somewhere around TYC\,8606-2025-1. 
This material should be ionized by UV photons from the white dwarf and therefore be visible at optical wavelengths.
The typical expansion velocity of PNe of $\approx25-30 \, \kms$ (Gesicki \& Zijlstra 2000) and the duration 
of the post-AGB phase of $\sim10\,000$\,yr imply that a hydrogen-rich nebula could be found at $\approx0.3$ pc 
from TYC\,8606-2025-1. Indeed, inspection of the SHS data led to the detection of an almost straight ($\approx45$ 
arcsec long) filament at about 90 arcsec (or $\approx0.42$ pc) to the northeast of TYC\,8606-2025-1 and an arcuate 
region of diffuse emission of angular radius of $\approx75$ arcsec (or $\approx0.35$ pc) to the southwest of 
TYC\,8606-2025-1, which apparently encircles the IR nebula (see Fig.\,\ref{fig:fil}). We speculate that these features 
are what remains of the slow AGB wind material swept-up by the fast post-AGB wind during the first post-AGB track. 
Unfortunately, the position angle of the RSS slit was chosen in such a way (see Section\,\ref{sec:rss}) that it did not
cross the optical features, which precludes us from making any decisive conclusion on their nature. Note that 
these features are not visible in the VPHAS+ H\,$\alpha$ image. This could be due to the larger FWHM of the VPHAS+ 
H\,$\alpha$ filter (107 \AA) as compared to that of SHS (70 \AA).

Finally, we note that the {\it Gaia} DR2 provides high-precision proper motion measurements for TYC\,8606-2025-1:
$\mu_\alpha =-3.55\pm0.06$ mas\,yr$^{-1}$ and $\mu_\delta \cos b =2.83\pm0.06$ mas\,yr$^{-1}$. After correction for 
the Galactic differential rotation and solar motion, one finds the peculiar transverse velocity of TYC\,8606-2025-1 to 
be $v_{\rm tr}=(v_l ^2+v_b ^2)^{1/2}=54.3\pm0.3 \, \kms$, where $v_l =46.8\pm0.3 \, \kms$ and $v_b =27.4\pm0.3 \, \kms$ 
are, respectively, the peculiar velocity components along the Galactic longitude and latitude. To calculate $v_{\rm tr}$, 
we used the Galactic constants $R_0 = 8.0$ kpc and $\Theta _0 =240 \, \kms$ (Reid et al. 2009) and the solar
peculiar motion $(U_{\odot},V_{\odot},W_{\odot})=(11.1,12.2,7.3) \, \kms$ (Sch\"onrich, Binney \& Dehnen 2010). 
For the error calculation, only the errors of the proper motion measurements were considered. The derived values of 
$v_l$ and $v_b$ indicate that TYC\,8606-2025-1 is a runaway star moving along a position angle of $\approx100$\degr,
i.e. almost from west to east. Surprisingly, there is no obvious correlation between the direction of the stellar motion
and the shape of the nebula. This could be understood if TYC\,8606-2025-1 is still embedded in the hydrogen envelope 
lost by the primary, which shields the nebula from space motion (cf. Gvaramadze et al. 2009).

\subsection{Further work}
\label{sec:fur}

Further work would be needed to check our conclusions about the nature of TYC\,8606-2025-1 and its 
associated circumstellar nebula(e). It includes:

\begin{itemize}

\item Search for UV and X-ray emission from TYC\,8606-2025-1 to confirm the presence of the white dwarf 
companion.

\item Narrowband imaging and spectroscopy of the optical features around TYC\,8606-2025-1 to establish their 
possible relationship with the star, and, potentially, to derive their chemical abundances. 

\item Constraining the orbital parameters of TYC\,8606-2025-1 using forthcoming {\it Gaia} DR3 data and new radial 
velocity measurements.

\item 3D hydrodynamic modelling of the spiral nebula.

\end{itemize}

\section{Summary}
\label{sec:sum}

We have discovered an IR spiral nebula using data from the {\it WISE} all-sky survey. Follow-up optical
spectroscopy with SALT showed that the star associated with the nebula, TYC\,8606-2025-1, is a G8\,III mild 
barium star. Spectral analysis of TYC\,8606-2025-1 along with the {\it Gaia} DR2 data allowed us to derive the 
luminosity of this star of $\log(L/\lsun)\approx2$ and its initial mass of $\approx3 \, \msun$. 
An analysis of our own and archival \'echelle spectroscopic data has revealed radial velocity variability in 
TYC\,8606-2025-1, meaning that this star is in a binary system and that its surface is polluted with barium 
(and other s-process elements) as a result of wind accretion from the more evolved companion star (now a 
white dwarf). We did not find an optical counterpart to the spiral nebula in the available sky surveys in 
spite of the low distance and reddening of the star, nor did we detect nebular lines in the long slit spectrum 
of the star. We have interpreted this non-detection as an indication that the nebula is composed of hydrogen-poor 
material shed by the companion (primary) star after the very late thermal pulse (VLTP), and that metals in this 
material are at least triply ionized and therefore do not form nebular lines at optical wavelengths. The very 
likely binary status of TYC\,8606-2025-1 implies that the formation of the spiral nebula around this star was 
induced by the orbital motion of the mass-losing post-VLTP star around the windless G star. From the shape and 
linear size of the nebula, we inferred that it was formed during approximately one orbital period of 
$\sim100$\,yr, which is of the same order of magnitude as the duration of the post-VLTP phase. Our suggestions 
on the origin of the IR nebula imply that it should be surrounded by a more extended optically visible nebula 
produced by a hydrogen-rich wind from the primary star during the AGB and early post-AGB phases. We have detected 
several optical filaments in the SHS H\,$\alpha$ +[N\,{\sc ii}] image of the region around the IR nebula. 
Unfortunately, our long slit spectroscopy did not cover these filaments, so that their possible connection to 
TYC\,8606-2025-1 remains unclear. 

\section{Acknowledgements}
This work is based on observations obtained with the Southern African Large Telescope (SALT), programmes 
2013-2-RSA$\_$OTH-003 and 2018-1-MLT-008, and collected at the European Southern Observatory under ESO programme 
095.A-9011(A). V.V.G. acknowledges support from the Russian Science Foundation under grant 19-12-00383 (analysis 
and interpretation of observational data) and from the Russian Foundation for Basic Research (RFBR) under grant 
19-02-00779 (search for infrared nebulae with {\it Spitzer} and {\it WISE}). A.Y.K. acknowledges support from 
RFBR under grant 19-02-00779 (spectroscopic observations and data reduction) and from the National Research 
Foundation (NRF) of South Africa. E.K.G. gratefully acknowledges funding by the Sonderforschungsbereich ``The 
Milky Way System" (SFB 881, especially subproject A5) of the German Research Foundation (DFG). This research 
has made use of the NASA/IPAC Infrared Science Archive, which is operated by the Jet Propulsion Laboratory, 
California Institute of Technology, under contract with the National Aeronautics and Space Administration,
the SIMBAD database and the VizieR catalogue access tool, both operated at CDS, Strasbourg, France, and data 
from the European Space Agency (ESA) mission {\it Gaia} (https://www.cosmos.esa.int/gaia), processed by the 
{\it Gaia} Data Processing and Analysis Consortium (DPAC, https://www.cosmos.esa.int/web/gaia/dpac/consortium). 
Funding for the DPAC has been provided by national institutions, in particular the institutions participating in 
the Gaia Multilateral Agreement.

\end{document}